\documentclass[a4paper]{article}

\usepackage{itgspeech2023}    
\usepackage{times}            
\usepackage[english]{babel}   
\usepackage[ansinew]{inputenc}
\usepackage[T1]{fontenc}      
\usepackage[sort&compress,numbers]{natbib}	
\usepackage{amsmath,amssymb}
\usepackage{graphicx}
\usepackage[colorlinks=false,pdfborder={0 0 0}]{hyperref}
\usepackage{units}
\usepackage{booktabs}
\usepackage{multirow}
\usepackage{amsmath,graphicx}
\usepackage{todonotes}
\usepackage{url}
\usepackage{verbatim}
\usepackage{balance}
\usepackage{colortbl}
 \usepackage{microtype}
 \usepackage[symbol]{footmisc}

 \usepackage[T1]{fontenc}
\title{Exploratory Evaluation of Speech Content Masking}

\author{Jennifer Williams $^1$, Karla Pizzi $^{2,3}$, Paul-Gauthier No\'e $^4$, Sneha Das $^5$}
\address{
$^1$ Electronics and Computer Science, University of Southampton, UK\\
  $^2$ Fraunhofer AISEC, Garching, Germany; $^3$ Technical University Munich, Germany\\
  $^4$ Laboratoire Informatique d'Avignon (LIA), University of Avignon, France\\
  $^5$ Dept. of Applied Mathematics and Computer Science, Technical University of Denmark, Denmark \\
Email: \texttt{j.williams@soton.ac.uk, karla.pizzi@aisec.fraunhofer.de}}

\begin{document}

\maketitle

\begin{abstract}
Most recent speech privacy efforts have focused on anonymizing acoustic speaker attributes but there has not been as much research into protecting information from speech content. We introduce a toy problem that explores an emerging type of privacy called ``content masking'' which conceals selected words and phrases in speech. In our efforts to define this problem space, we evaluate an introductory baseline masking technique based on modifying sequences of discrete phone representations (phone codes) produced from a pre-trained vector-quantized variational autoencoder (VQ-VAE) and re-synthesized using WaveRNN. We investigate three different masking locations and three types of masking strategies: noise substitution, word deletion, and phone sequence reversal. Our work attempts to characterize how masking affects two downstream tasks: automatic speech recognition (ASR) and automatic speaker verification (ASV). We observe how the different masks types and locations impact these downstream tasks and discuss how these issues may influence privacy goals. 
\end{abstract}


\section{Introduction}
Audio recordings comprise private information in many ways.
In this paper, we focus on content-related information, ranging from cues about location or acoustic scene~\cite{tang2020scene} to sensitive words and phrases. 
Speech data can therefore expose privacy risks even if a speaker's voice has been anonymized, since typical goals for voice anonymization include preserving intelligibility \cite{tomashenko2022voiceprivacy}. 
Content involving words or phrases could be used (or misused) to reveal personal data such as birth date or credit card details, or enable an attacker to identify a person based on spoken names or other similar references.
Therefore, the capability to achieve fine-grained control over masking particular aspects of speech data beyond speaker voice characteristics must become center-stage for privacy protection research.

In this paper, we introduce a toy problem to explore the concept of content privacy for speech, as a complement to the extensive ongoing efforts from the biennial Voice Privacy Challenges\footnote{\url{https://www.voiceprivacychallenge.org/}.} (VPC). \color{black}
Systems developed through the VPC aim to conceal, remove, or neutralize speaker-related voice attributes. 
While the VPC systems have many practical uses, that singular form of privacy is not enough to achieve full anonymization, due to the ability to link spoken content.
Consider the example of an audio recording in a dataset dealing with medical health issues with the following utterance: \textit{``My name is James Smith and I have severe depression.''} 
Since the speaker reveals information in the spoken content, anonymizing the acoustic-related speaker attributes alone would not conceal the sensitive information (the person's identity and health condition) since this can be obtained simply by analyzing the words. 
Sharing speech data across research areas (e.g., from the medical research domain to the speech recognition domain) is difficult due to compliance with GDPR \cite{nautsch2019gdpr} and other related ethical concerns. 
Useful applications of content masking include sharing sensitive speech databases for research purposes, concealing sensitive words or phrases from speech synthesis screen readers, and disentanglement of speech features and content for more deeper privacy controls. 
If there was a way to achieve 
full anonymisation and redact all sensitive information reliably, this could make it more feasible to share datasets that are important for speech research, ultimately benefiting speech technology development. While targeted aspects of the speech signal become incomprehensible from masking, the speech signal should only be `destroyed' on private information. Our approach serves as foundational work, presenting a first-ever baseline approach using speech re-synthesis. 


Since we are introducing content privacy and its evaluation in this paper, there are no established techniques or evaluation protocols \cite{williams2021revisiting,williams2022challenges}. 
We adopt a re-synthesis technique based on vector-quantized variational autoencoders (VQ-VAE) \cite{van2017neural} and present baselines for masking content. 
We do not propose that VQ-VAE is the best content masking approach, but instead we assess whether re-synthesis techniques could be useful. \color{black}
The VQ phone codes \cite{tjandra2019vqvae} are sequential and preserve temporal structure of speech, meaning that a string of phones is temporally related to the sequence of words \cite{fong2021analysing}. 
To mask target words in the speech, we modify sub-strings of the VQ sequence and re-synthesize from the VQ codes using WaveRNN \cite{paul2020speaker}. 
In this paper, we explore masking using different types of masks (noise, deletion, and reversal), and we target words at different locations in the utterances (start, middle, and end). 
We evaluate masking using two downstream tasks: automatic speech recognition (ASR), and automatic speaker verification (ASV). 
We investigated ASR and ASV performance to explore how this type of problem could be evaluated.  

We also perform masking on original natural speech by removing or replacing segments corresponding to the three types of mask. 
This baseline simulates perfect masking with perfect quality of speech, to compare against the VQ-VAE speech re-synthesis technique. The contributions of this paper are: 
\begin{itemize}
    \item Explore how to modify and replace sequences of VQ-VAE phone codes to conceal content of selected words and phrases
    \item Understand how masking words in an utterance will affect downstream tasks such as ASR and ASV by masking original speech (assumes perfect synthesis) as well as re-synthesis from VQ-VAE
\end{itemize}

\section{Related Work}

Previous work in masking speech content involved concealing conversations that take place indoors using a technique called \textit{scrambling}~\cite{de2008speech}. 
This technique was reported to be very effective at masking speech in order to make it unavailable to eavesdroppers~\cite{bopardikar2005speech,ma1996wavelet, jayant1981comparison}. 
One of the issues is that the method depends on knowing the room characteristics beforehand. 
The method does not generalise to other use-cases wherein we may want a finer-grained approach to conceal content without disturbing neighboring words or while preserving speaker voice characteristics. 

Recent research has explored content privacy for ASR, specifically for training language models on texts that have had named entities redacted due to privacy~\cite{turan2022adapting}. 
Their redaction technique (called \textit{sanitization}) involved swapping one named entity for another. 
They found that the technique results in significant increase in word error rate (WER) when training new language models. 
The drop in performance is likely due to the sequential nature of language modeling.
The authors did not explore other types of sanitization.
With this paper, we aim at filling this gap.
We use a method presented in \cite{williams21_ssw}.
Here, content masking using VQ-VAE was mentioned as being a possible technique for masking. 
However, their results are difficult to interpret because the evaluation involved human judgments of speaker similarity and WER between a very small sample of original and masked utterances.   

A privacy-preserving ASR system, called Pr$\varepsilon\varepsilon$ch~\cite{ahmed2020preech}, was recently proposed to preserve voice biometrics and content privacy for offline and cloud-based ASR services. 
Specifically, the system employs a combination of segmentation and shuffling of signals, identification and removal of sensitive information, and injection of dummy phrases to preserve content privacy. 
Unfortunately, the data is transformed into a `bag of words' which may make it less useful for further downstream applications outside of ASR.

Beyond this, adversarial techniques for speech information hiding have been recently proposed in \cite{dong2022adversarial}, which demonstrates a method to encode utterances into a database and recover them later using a neural network. 
The approach relies heavily on forced alignments. 
Further, they do not assess impacts on downstream tasks nor do they conceal words/phrases at specific positions in an utterance. Whether private information should be recoverable depends upon the use-case. In our work, we consider both cases where recovery is and is not desirable. Recovery is especially undesirable for sensitive private information due to adversarial attacks. 

\section{Masking Methodology}

\subsection{Data and Pre-Processing}
The data that we use in this paper was the Voice Cloning Toolkit (VCTK) v0.92 \cite{yamagishi2019cstr}. 
The dataset contains 110 unique speakers, both male and female, of varying ages and English accents. 
We chose the VCTK dataset because it is studio-quality speech, which helps us establish our content masking baselines. 
We obtained forced alignments for each audio file from the Montreal forced aligner  \cite{mcauliffe2017montreal}. 
As this paper explores the concept of content masking, we made use of forced alignments to ensure that we were able to target specific words at locations in each utterance. 
In real-world applications of content masking, it may not be possible or desireable to use forced alignments, and instead a technique such as keyword spotting would be more appropriate depending on the specific application scenario.   

\subsection{Masking Technique}
\label{sec:mask_technique}
We compare masking for two types of speech: original (natural) utterances and re-synthesized utterances. 
Words are masked in original speech by relying on forced alignments. 
This provides us with high-quality masked speech, and sets an upper-bound as if the speech re-synthesis and learned latent representations were of perfect quality. 
We are interested in VQ-VAE speech re-synthesis because this technique can learn rich latent representations using autoencoders. 
For our baselines involving natural speech from VCTK, we replace segments of natural speech with the masks. 
For baselines involving speech re-synthesis with VQ-VAE, we modify sequences of VQ phone codes that representing the target words to be masked.

One of the benefits of using VQ-VAE is that it allows us to consider content masking while speech is in a coded or compressed state \cite{casebeer2021enhancing}, wherein the masking is done strictly by manipulating sequences of VQ phone codes.
This mimics realistic scenarios where speech may be compressed or in transmission and we may have some temporal information or keys to designate which sequences of codes to conceal for privacy purposes. 
The VQ-VAE system that we used in this paper comes from \cite{williams2021learning}.
It is a dual-encoder model which models phone content information and speaker identity. 
Using that system and pre-trained model, we explored three different locations of the mask within an utterance, and three types of masks.

\subsection{Mask Types}
Many types of masks are relevant to speech content masking. 
For example, it is common in live television to replace certain offensive words with a long `beep' sound \cite{williams2021revisiting}. 
In this work, we examine three types of masks and describe them here.
Note that all mask types can change the usual structure of the sentence, which might potentially lead to wrong transcriptions if an automatic speech recognition (ASR) system with a language model is in place.

\textbf{Noise.} 
The noise mask involves replacing content with a temporally-modulated speech-shaped noise masker (ICRA noise 9 from~\cite{cooke2013evaluating}). 
This was proposed by \cite{williams2021learning}, however they did not evaluate how masking affects downstream tasks. 
When masking original speech, we use forced alignments as a guide to replace segments of the natural speech with an identical length segment of the speech-shaped noise masker. When masking re-synthesized speech using VQ codes, we first obtain VQ codes corresponding to noise-only and then swap the sequence of VQ codes (of the same length). 

\textbf{Deletion.} 
A deletion mask involves deleting content entirely from an utterance and the content will not be recoverable. 
For natural speech, we slice the waveform and remove the segment that contains the target words. 
For re-synthesis, we remove the VQ phone code sequence corresponding to the target words, before synthesis. 
We include it in this paper in order to assess impacts on speech re-synthesis as well as impacts on the downstream tasks.

\textbf{Reversal.} 
We explore a third mask type that reverses target words in the time-domain. 
For original speech, this involves slicing the waveform and reversing only the segment that corresponds to the target words. 
For re-synthesis, we simply reverse the sequence of VQ codes while keeping the surrounding VQ code sequences in-tact. 


\subsection{Materials for Evaluation}
We compiled a subset of speech samples with and without masking. 
The speakers and utterances were selected from the ``condition \#3'' that was reported by the authors who provided the system and pre-trained model \cite{williams2021learning}. 
This particular condition contained samples wherein the speakers were unseen during training and the content was seen during training. 
We selected this subset since we are especially interested in content masking, but wanted to avoid using data that had been fully seen during training of the VQ-VAE system. 
From this subset, we selected 9 speakers (\textit{p260}, \textit{p285}, \textit{p294}, \textit{p300}, \textit{p305}, \textit{p307}, \textit{p310}, \textit{p347}, and \textit{p351}). 
These speakers were selected because they had an average speaking rate of $<$ 5 words/second, the total duration of the utterances contained~$>$~300 VQ phone codes (1.2 seconds), and number of words spoken was $\ge$ 7. 
When counting words, we ignored all ``SIL'' tokens that were generated from forced alignments by the Montreal aligner. 
With these speakers and utterances, we applied the masking to create samples for downstream evaluation\footnote{\url{https://rhoposit.github.io/ITG2023}}.

\section{Masking Evaluation}
\subsection{Automatic Speech Recognition (ASR)}
To measure the effect of masking on speech recognition, we employed four state-of-the-art ASR models: two SpeechBrain systems \cite{speechbrain} and two Whisper systems \cite{radford2022robust}.
The two SpeechBrain systems are sequence to sequence-based models with a combination of convolutional, recurrent, and fully-connected networks (CRDNN) trained on LibriSpeech \cite{panayotov2015librispeech}:
(1) a CRDNN with an RNN-based language model (\textbf{SB-RNN})\footnote{ \url{https://huggingface.co/speechbrain/asr-crdnn-rnnlm-librispeech}}; and
(2) a CRDNN with a transformer-based language model (\textbf{SB-Transformer})\footnote{ \url{https://huggingface.co/speechbrain/asr-crdnn-transformerlm-librispeech}}.
For Whisper, we used the ``base'' (\textbf{WH-Base}) as well as the ``medium'' (\textbf{WH-Medium}) models.
Both Whisper models are end-to-end encoder-decoder transformer models.
Due to the architecture and the variety of data Whisper has been trained on, it exhibits improved robustness to accents, background noise and technical language \cite{radford2022robust}.

\begin{table}[ht]
    \small
    \centering
    \begin{tabular}{l||c||c|c|c}
    \hline
        & Original & \multicolumn{3}{c}{Original Speech + Masking}\\   
        System &  Speech & Noise & Deletion & Reversal\\  \hline
        SB-RNN & 13.9 & 22.9 & 20.6 & 37.5 \\
        SB-Trans. & 10.5 & 19.6 & 17.2 & 36.4\\ 
        WH-Base & 4.15 & 10.6 & 10.5 & 32.2 \\
        \textbf{WH-Medium} & \textbf{1.24} & \textbf{9.30} & \textbf{7.72} & \textbf{24.7}\\
        \hline \hline
        & VQ-VAE & \multicolumn{3}{c}{VQ-VAE + Masking}\\   
        System &  Speech & Noise & Deletion & Reversal\\
        \hline
        SB-RNN & 47.5 & 75.5 & 65.2 & 78.3 \\
        SB-Trans. & 38.8 & 76.0 & 59.0 & 75.4 \\
        WH-Base & 47.7 & 78.7 & 66.2 & 79.5\\
        \textbf{WH-Medium} & \textbf{32.3} & \textbf{63.0} & \textbf{55.3} & \textbf{69.1} \\
        \hline
    \end{tabular}
    \caption{ASR system $\downarrow$WER\% for original and re-synthesised un-masked and masked utterances.}
    \label{tab:ASR_results}
\end{table}

\begin{figure}[ht]
\centering
\includegraphics[width=8.5cm]{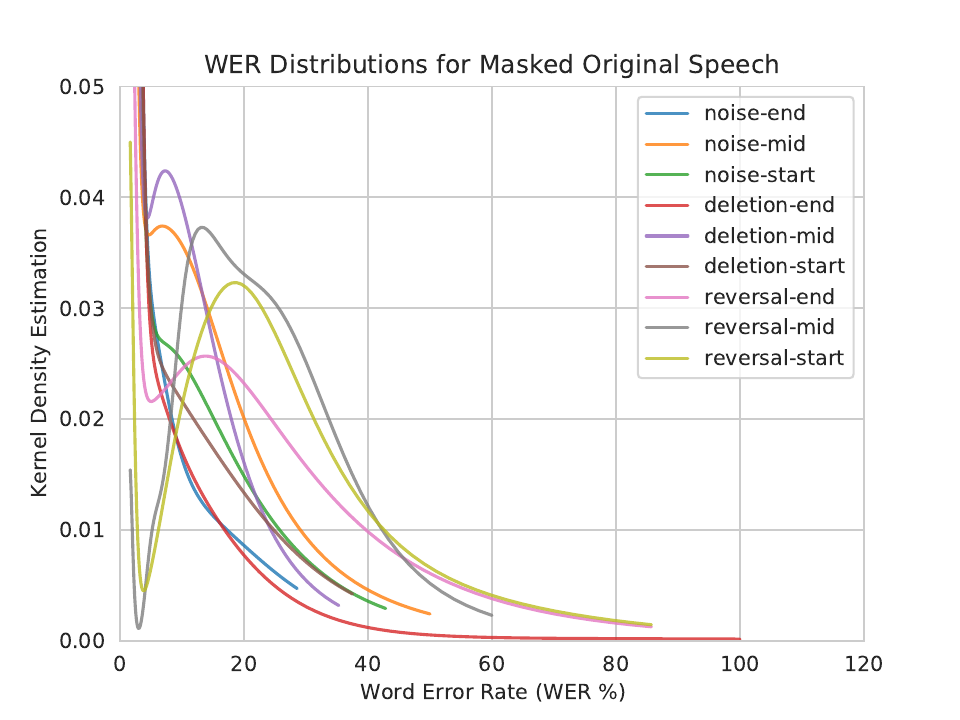}
\caption{Kernel density estimation of $\downarrow$WER\% for \textbf{WH-Medium} on original speech that was masked. This plot shows the mask types and positions (averaged across all speakers). }
\label{fig:asr_wer_original_masked_position}
\end{figure}
\vspace{-1mm}

We examine the impact of masking on ASR by measuring word error rate (WER\%) while comparing the four different ASR models, as shown in Table~\ref{tab:ASR_results}. 
First, we evaluated each ASR system using un-masked speech from the original and VQ-VAE re-synthesis to establish a benchmark of performance on the VCTK dataset, using the original VCTK transcripts as reference. 
Next, we evaluated ASR on masked speech using the masking techniques described in Section~\ref{sec:mask_technique}, using VCTK transcripts that had the target (masked) words removed, as the reference.
In some cases, ASR systems demonstrated catastrophic failure, such as repeating words 5 or more times, which occurred most often with the noise and reversal masks. 
We omitted utterances wherein the ASR transcription was more than 30 characters longer than the reference transcript. Even with this quality control in place, we did occasionally observe WER values above 100\%, but less so with the \textbf{WH-Medium} model.

\begin{table*}[ht]
    \centering
    \begin{tabular}{l||c||ccc|ccc|ccc}
    \hline
    & \textbf{None} & \multicolumn{9}{c}{\textbf{With Masking}} \\
    \hline
     Mask position & \multirow{2}{*}{--} & \multicolumn{3}{c|}{Start} & \multicolumn{3}{c|}{Middle} & \multicolumn{3}{c}{End} \\ 
     Mask type & & Noise & Delete & Reversal & Noise & Delete & Reversal & Noise & Deletion & Reversal \\
     \hline
    Original & 0.04 & 0.02 & 0.08 & 0.13 & 0.27 & 0.07 & 0.17 & 0.07 & 0.00 & 0.00 \\
    VQ-VAE & 16.75 & 19.65 & 22.53 & 23.35 & 21.81 & 19.85 & 23.13 & 23.89 & 22.29 & 21.80 \\
    \hline
    \end{tabular}
    \caption{Comparison of ASV performance measuring $\downarrow$EER\% using original un-masked enrollment utterances and either original or VQ-VAE masked test utterances. }
    \label{tab:ASV_results}
    \vspace{-4mm}
\end{table*}

The WER for un-masked audio and un-masked transcription represents our empirical WER. 
We conducted a paired t-test ($p\ll0.05$, 95\% confidence) for each ASR system separately, and found that the results are significant except for original masked speech, where noise and deletion masks are not statistically significant. 
Likewise, the differences between noise and reversal are not significant for VQ-VAE masked speech for the SpeechBrain models. 
For \textbf{SB-RNN} and \textbf{SB-Transformer}, these perform worse overall compared to the Whisper models. 
This is likely due to SpeechBrain models being trained on LibriSpeech, whereas Whisper models are trained on a diverse dataset, including different English dialects. 
All models indicated worse performance when reversal masking was used. 
Issues with reversal masking may result from audio retaining speech features, as tonal and atonal sounds are roughly kept, while the ASR system is not able to detect meaningful words.
This is especially problematic if a language model is in place.

\begin{figure}[ht]
\centering
\includegraphics[width=8.5cm]{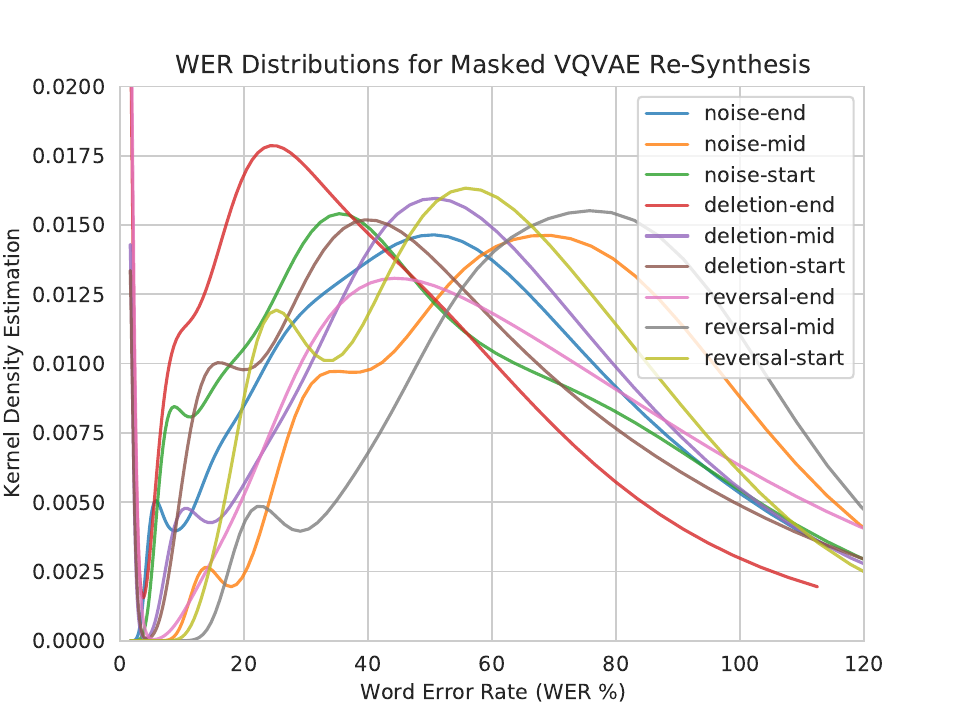}
\caption{Kernel density estimation of $\downarrow$WER\% for \textbf{WH-Medium} on VQ-VAE re-synthesized speech that was masked. This plot shows the mask types and positions (averaged across all speakers). }
\label{fig:asr_wer_vqvae_masked_position}
\end{figure}

From these initial results, we further characterized ASR evaluation using only \textbf{WH-Medium} since it performed best on our original and VQ-VAE speech. 
We show the kernel density estimations (KDE)\footnote{Gaussian KDE has an unbounded support $]-\infty,+\infty[$. 
Because the WER is positive, we estimate the density of the log WER. 
The estimated density is then mapped back in the WER domain using the change of variable formula and nomalization. As it is a density estimation (a curve fit to a histogram) there is typical congestion at zero.} of the distributions of per-utterance WER scores in Figure~\ref{fig:asr_wer_original_masked_position} and Figure~\ref{fig:asr_wer_vqvae_masked_position}. 
These two figures are showing that certain mask positions and types result in different WER. 
In the original masked speech (Fig.~\ref{fig:asr_wer_original_masked_position}), the noise mask at the end of an utterance had the least negative impact on ASR performance whereas the reversal at the middle of an utterance generally had a large impact. 
%
Performance on VQ-VAE speech for the \textbf{WH-medium} model (Fig.~\ref{fig:asr_wer_vqvae_masked_position}) also shows that WER differs by mask location.
The highest WERs came from masking words in the middle of a sentence.
This may be due to the ASR model architectures, as the middle influences both previous and subsequent word transcriptions, or due to speech re-synthesis which used a recurrent network in the WaveRNN vocoder. 


\vspace{-2mm}


\subsection{Automatic Speaker Verification (ASV)}

In the ASV evaluation, we compare each utterance with all others, resulting in 460 \emph{target} trials and 4096 \emph{non-target} trials. The cosine similarity between the ECAPA-TDNN speaker embedding~\cite{ecapatdnn} of each utterance is used as the comparison score. 
Table~\ref{tab:ASV_results} shows the results in terms of equal-error rate (EER\%). 
EER designates the error rate where both the false acceptance and false rejection rate are equal. 
Lower EER scores indicate better discrimination performance. 
Because the number of speakers being tested is small, it is therefore delicate to draw general conclusions. 
However, we can already see from the first line (original speech), a slight drop of performance when the mask is applied in the middle of the sample with both noise and reversal type on original speech. 
This could be due to the fact that the attention mechanism of ECAPA-TDNN, which has a similar role as a voice activity detector, does not properly ignore the mask when it is at the middle of an utterance. 
For the deletion mask, we do not observe this issue because the segment being protected is cut from the audio, so there is therefore nothing for the attention to try and ignore. 
From the second line (VQ-VAE speech), we can see that, without masking, applying the VQ-VAE significantly perturbs the speaker information with an increase of EER from 0.04 to 16.75. 
Applying the masking further degrades the ASV performance but without significant differences between mask type and position. 
In future works, analysing how the attention of the ECAPA-TDNN behaves against the mask could help in better understanding this drop of performance.




\section{Summary and Future Work}
We have presented a framework and evaluation for masking speech content that uses both original and re-synthesised utterances.
While masking natural speech represents an idealized baseline, the re-synthesized speech has been compressed into a rich latent space using VQ-VAE.
We assessed how masking the utterance content can affect performance with ASR and ASV as two relevant downstream tasks. 
We showed that mask types and locations affect ASR and ASV performance differently.
\color{black}


Our chosen toy problem masking technique allowed us to work directly with audio-only data, but the RNN-based vocoder suffers from modified VQ sequences as evidenced by poor WER performance mid-utterance. We therefore propose that future experiments explore a variety of vocoders beyond WaveRNN.
We were not able to compare impacts of masking across genders due to the small amount of data in our concept problem but we plan to investigate this in future work. Further, datasets other than VCTK may be affected by masking differently. 
Importantly, both the ASV and ASR tasks are impacted by the speech re-synthesis quality. We expect the field of speech synthesis to continually improve. Our idea of content masking is based on speech compression and re-synthesis that can tolerate mid-utterance disruptions, similar to disfluencies or signal drop-outs. 
For real-world applications, we are interested to further explore masking methods that allow for recovery of the original content, including the role of an adversary. \color{black}
We are also interested to conduct future work that can address whether particular types of masks are more or less difficult for an attacker to discover the hidden content, including what kind of information an adversary would need in order to reconstruct hidden content. 
We would like to find out if we can trick a speech or speaker recognition system on purpose, based on content masking techniques.

\section*{Acknowledgements}
This work was partially supported by the UK EPSRC Trustworthy Autonomous Systems Hub (EP/V00784X/1); a JST CREST Grant (JPMJCR18A6, VoicePersonae project), Japan; the Bavarian Ministry of Economic Affairs, Regional Development, and Energy as well as the German Federal Ministry for the Environment, Nature Conservation, Nuclear Safety and Consumer Protection.




\small
\balance
\bibliographystyle{ieeetr}
\bibliography{mybib}


\end{document}